\documentclass[journal]{IEEEtran}

\ifCLASSINFOpdf

\else

\fi
\usepackage{algorithm}      
\usepackage{algpseudocode}  
\usepackage{amsfonts}
\usepackage{amssymb}
\usepackage{pgfplots}
\usepackage{tikz}
\usepackage{pgfplots}
\pgfplotsset{compat=1.5}
\usepackage{graphicx}
\usepackage{subcaption}
\usepackage{amsmath}
\usepackage{dsfont}
\usepackage{graphicx}
\usepackage{cite}
\usepackage{bigstrut}
\usepackage{float}
\usepackage{balance}
\usepackage{lipsum}
\usepackage{hyperref}
\usepackage{psfrag}
\usepackage{xcolor}
\usepackage{tikz}
\usepackage{enumitem}
\usepackage{xcolor}
\usepackage{pgfplots}
\usepackage{physics}
\usepackage{amsmath}
\usepackage{tikz}
\usepackage{mathdots}
\usepackage{yhmath}
\usepackage{cancel}
\usepackage{color}
\usepackage{siunitx}
\usepackage{array}
\usepackage{multirow}
\usepackage{amssymb}
\usepackage{gensymb}
\usepackage{tabularx}
\usepackage{extarrows}
\usepackage{booktabs}
\usetikzlibrary{fadings}
\usetikzlibrary{patterns}
\usetikzlibrary{shadows.blur}
\usetikzlibrary{shapes}

\begin{document}

\title{OFDMA for Pinching-Antenna Systems}

\author{Thrassos K. Oikonomou,~\IEEEmembership{Graduate Student Member,~IEEE}, Sotiris A. Tegos,~\IEEEmembership{Senior Member,~IEEE}, \\ Panagiotis D. Diamantoulakis,~\IEEEmembership{Senior Member,~IEEE}, Yuanwei Liu,~\IEEEmembership{Fellow,~IEEE},\\ and George K. Karagiannidis,~\IEEEmembership{Fellow,~IEEE}
\thanks{T. K. Oikonomou, S. A. Tegos, P. D. Diamantoulakis and G. K. Karagiannidis are with the Department of Electrical and Computer Engineering, Aristotle University of Thessaloniki, 54124 Thessaloniki, Greece (e-mails: toikonom@auth.gr, tegosoti@auth.gr, padiaman@auth.gr, geokarag@auth.gr).}
\thanks{Yuanwei Liu is with the Department of Electrical and Electronic Engineering, The University of Hong Kong, Hong Kong (e-mail: yuanwei@hku.hk).}
\vspace{-4mm}
}

\maketitle
\begin{abstract}
Pinching-antenna (PA) systems route millimeter wave (mmWave) signals through a leaky waveguide and radiate them at ``pinch" apertures, offering low-cost line-of-sight (LoS) coverage. However, when multiple PAs serve multiple users simultaneously, the downlink channel becomes strongly frequency-selective, creating inter-symbol interference (ISI) that existing single-carrier designs overlook. This paper models the overall channel as a finite impulse response (FIR) filter, characterizes its frequency selectivity, and explicitly accounts for the resulting ISI. To overcome ISI, we introduce an orthogonal frequency-division multiple access (OFDMA)-based framework and formulate a max-min resource-allocation problem to achieve user fairness. A lightweight two-stage heuristic-greedy subcarrier assignment, followed by per-user water-filling, achieves near-optimal fairness with polynomial complexity. Simulation results for an indoor layout demonstrate that the proposed scheme notably increases the minimum user rate compared to time-division single-carrier baselines and remains robust under moderate LoS blockage.
\end{abstract}


\section{Introduction}
Achieving multi-gigabit throughput in next-generation wireless systems requires highly directional transmission methods that avoid complex hardware, leading research into millimeter-wave (mmWave) links, large-scale antenna arrays, and reconfigurable intelligent surfaces \cite{yuanwei}. In this direction, pinching antenna (PA) systems have recently been proposed to route radio-frequency energy through a leaky dielectric waveguide and introduce small “pinch” apertures at selected positions, allowing each aperture to radiate \cite{docomo,karag_magazine,liu2025pinchingantennasystemspassarchitecture, yuanwei_beamforming}. Since the signal remains guided inside the waveguide until it reaches the pinch point, this configuration reduces free-space path loss while maintaining a cable-like footprint. Activating pinch points at specific locations enables the base station (BS) to establish line-of-sight (LoS) links with multiple indoor terminals, minimizing propagation loss and establishing the PA technology as a compact, cost-effective solution for future high-frequency deployments.

Early research has explored several facets of this new architecture. In \cite{Ding1}, the authors quantified the path loss reduction and steering agility that adaptive PA positioning can achieve, while in \cite{tyr}, closed-form expression for coverage probability of randomly located users were derived. Furthermore, the downlink performance of a fixed PA layout was analyzed in \cite{ding_karag_downlink}, revealing substantial rate gains over free-space baselines when the LoS path is maintained. In multi-user scenarios, the authors of \cite{sotiris_panos} proposed a scheduling strategy that balances user rates, while in \cite{ding_blockage}, it was shown that certain LoS blockages can increase sum throughput by decorrelating users. However, all of these studies employ single-carrier transmission and mitigate intersymbol interference (ISI) by dynamically clustering pinch apertures directly above the scheduled user, keeping their spacing small enough that the associated delay spread becomes negligible. Consequently, none of these works models the downlink as a finite impulse response (FIR) filter or quantify the frequency selectivity that naturally arises when several pre-placed pinches serve multiple users simultaneously. In such practical, low-complexity deployments, where PAs are fixed along the waveguide to provide concurrent LoS connectivity without continuous tracking, each terminal receives multiple delayed replicas of every symbol, and the resulting ISI becomes the main obstacle to high data rates. To the best of the authors’ knowledge, this intrinsic ISI has never been analyzed, nor has an OFDM-based solution been proposed for PA systems, which is the main motivation of this work.

Considering the above, we introduce a comprehensive physical-layer framework for PA systems that models the overall channel as an FIR filter, characterizes its frequency selectivity, and explicitly accounts for the resulting ISI. Leveraging this model, we formulate a max-min orthogonal frequency-division multiple access (OFDMA) resource-allocation problem that maximizes the worst-user rate under realistic power and orthogonality constraints and achieves user fairness. A practical two-stage heuristic-greedy subcarrier assignment, followed by per-user water-filling, utilizes the channel’s inherent frequency diversity, suppresses ISI without relocating the PAs, and retains polynomial complexity in the number of subcarriers. Simulation results demonstrate that the proposed scheme enhances minimum user throughput, outperforming time-division single-carrier benchmarks and expanding the operational capacity of PA networks in dense, high-capacity wireless environments.

\section{System Model}

We consider a downlink communication scenario consisting of a BS with $N$ PAs in a set $\mathcal{N}=\{1,\dots, N\}$ and $M$ single-antenna users in a set $\mathcal{M}=\{1,\dots, M\}$. Considering a three-dimensional coordinate system, we assume that the users are randomly deployed in a rectangular area in the x-y plane with sides $D_x$ and $D_y$, and $\boldsymbol{\psi}_m = (x_m, y_m,0)$ denotes the position of the $m$-th user. Specifically, $x_m$ follows a uniform distribution in $[0,D_x]$ and $y_m$ follows a uniform distribution in $[-D_y/2, D_y/2]$.

In the considered PA system, without loss of generality, the waveguide is assumed to be installed parallel to the x-axis with a height of $d$ and a length of $D_x$, which corresponds to one of the sides of the rectangular deployment area. To reduce system complexity, path loss, and the risk of outage due to blockage, the PAs are uniformly deployed along the waveguide. The position of the $n$-th PA is given by $\boldsymbol{\psi}_n^P = \left(x_n^P, 0, d\right)$, where $x_n = \frac{n D_x}{N+1}$. The channel between the $m$-th user and the $n$-th PA is given by
\begin{equation}
    \begin{aligned}
        h_{m,n} =  \frac{\alpha_{m,n}\sqrt{\eta}e^{-j\frac{2\pi}{\lambda}\left\|\boldsymbol{\psi}_m-\boldsymbol{\psi}_{n}^P\right\|}}{\left\|\boldsymbol{\psi}_m-\boldsymbol{\psi}_{n}^P\right\|},
    \end{aligned}
\end{equation}
where $\eta = c^2/(16\pi^2f_c^2)$ is a constant with $c$, $f_c$, and $\lambda$ denoting the speed of light, the carrier frequency, and the wavelength in free space, respectively. 
Furthermore, $\alpha_{m,n}$ is a binary indicator function representing the presence or absence of a LoS link between the $n$-th PA and the $m$-th user. Specifically, $\alpha_{m,n}=1$ if the LoS path is unblocked, and $\alpha_{m,n}=0$ otherwise \cite{ding_blockage}. The probability of an unobstructed LoS link between the $m$-th user and the $n$-th PA is modeled as \cite{blockage_modeling}
\begin{equation}
    \begin{aligned}
        \mathbb{P}(\alpha_{m,n} =1) = e^{-\beta\left\|\boldsymbol{\psi}_m-\boldsymbol{\psi}_{n}^P\right\|},
    \end{aligned}
\end{equation}
where $\beta$ is the LoS blockage parameter that characterizes the density and distribution of potential obstacles in the environment. Typical values for $\beta$ range from $0.05$ in sparsely furnished or open indoor spaces to $0.15$ in more cluttered indoor settings \cite{ding_blockage}. Since all $N$ PAs are placed along the same waveguide structure, the signal radiated by each PA is a phase-shifted replica of the signal fed into the waveguide at the BS. As a result, the equivalent channel through the waveguide can be expressed as
\begin{equation}
    \begin{aligned}
        h_{0,n}=e^{-j\frac{2\pi}{\lambda_g}\left\|\boldsymbol{\psi}_0-\boldsymbol{\psi}_{n}^P\right\|},
    \end{aligned}
\end{equation}
where $\lambda_g=\frac{\lambda}{n_e}$ is the guided wavelength with $n_e$ being the effective refractive index of the dielectric waveguide. Although $n_e$ varies with frequency due to waveguide dispersion and cutoff effects, we can treat it as constant across the bandwidth of interest if we operate well above the cutoff frequency, avoiding the need to model its frequency dependence.
Moreover, the BS transmits under a total power budget $P_t$, which is distributed across the $N$ PAs and across the available time-frequency resources. Finally, we consider that the users are served orthogonally.

\section{Proposed Approach}
In the downlink PA system, each user receives multiple delayed replicas of the transmitted symbol, with one arriving from each PA. Previous studies have mainly focused on how these delays affect the amplitude of the received signal, since the replicas can combine constructively or destructively. However, when the PAs are positioned far enough apart, the resulting propagation delays may extend over multiple symbol periods. This can lead to severe ISI. Under these conditions, the overall channel response can be accurately modeled as an FIR filter, given by
\begin{equation}
\label{eq:time_response}
    h_m(\tau)=\sum_{n=1}^N h_{0,n}h_{m,n}\delta\left(\tau-\tfrac{\left\|\boldsymbol{\psi}_m-\boldsymbol{\psi}_{n}^P\right\|}{c}-\tfrac{n_e\left\|\boldsymbol{\psi}_0-\boldsymbol{\psi}_{n}^P\right\|}{c}\right),
\end{equation}
where the Dirac impulse located at each composite delay accounts for the free-space propagation from the $n$-th PA to user $m$ and for the propagation from the feed point of the waveguide to this PA. Let the baseband waveform intended for the $m$-th user be expressed as a pulse train given by
\begin{equation}
    \begin{aligned}
        x_m(t) = \sum_{l=-\infty}^{\infty} s_{m,l}g(t-lT),
    \end{aligned}
\end{equation}
where $s_{m,l}$ is the $l$-th information symbol, $g(t)$ is a pulse of duration $T = 1/B$ with $B$ being the bandwidth of the system. Passing $x_m(t)$ through the FIR channel $h_m(\tau)$ in \eqref{eq:time_response} yields the time-domain observation, which is written as
\begin{equation} \label{eq::convolution}
    \begin{aligned}
        y_m(T) = h_m(t)*x_m(t) + w_m(t),
    \end{aligned}
\end{equation}
where $*$ denotes convolution and $w_m(t)$ represents the additive white Gaussian noise (AWGN) with zero mean and variance $\sigma^2$. It can be observed from \eqref{eq::convolution} that each symbol is convolved with multiple delayed replicas of the transmitted signal, thus the ISI is introduced by the spatially separated PAs.
In this context, taking the Fourier transform of \eqref{eq:time_response} yields the passband frequency response of the channel, which is given by
\begin{equation}
\label{eq:frequency_response}
H_m(f)=\sum_{n=1}^N h_{0, n}h_{m, n}   e^{-j 2 \pi\left(f-f_c\right) \left(\frac{\left\|\boldsymbol{\psi}_m-\boldsymbol{\psi}_{n}^P\right\|}{\lambda f_c}+\frac{\left\|\boldsymbol{\psi}_0-\boldsymbol{\psi}_{n}^P\right\|}{\lambda_g f_c}\right)},
\end{equation}
thus, in the frequency domain, the received signal can be expressed as 
\begin{equation}
    \begin{aligned}
        Y_m(f) = H_m(f)X_m(f) + W_m(f),
    \end{aligned}
\end{equation}
where $X_m(f)$ and $W_m(f)$ denoting the Fourier transforms of $x_m$ and $w_m$, respectively.

\begin{figure}[h]
\centering
\begin{tikzpicture}[x=0.75pt,y=0.75pt,yscale=-0.80,xscale=0.8]
\draw   (272.97,56) -- (517,56) -- (412.41,232.6) -- (168.38,232.6) -- cycle ;
\draw  [dash pattern={on 4.5pt off 4.5pt}]  (168.38,232.6) -- (298.97,11.58) ;
\draw [shift={(300.5,9)}, rotate = 120.58] [fill={rgb, 255:red, 0; green, 0; blue, 0 }  ][line width=0.08]  [draw opacity=0] (8.93,-4.29) -- (0,0) -- (8.93,4.29) -- cycle    ;
\draw  [fill={rgb, 255:red, 235; green, 232; blue, 233 }  ,fill opacity=1 ] (486.5,92) -- (219.47,92.7) -- (219.5,104) -- (486.53,103.3) -- cycle ;
\draw  [fill={rgb, 255:red, 74; green, 74; blue, 74 }  ,fill opacity=1 ] (270,90) -- (285.5,90) -- (285.5,105.5) -- (270,105.5) -- cycle ;
\draw (394.75,208) node  {\includegraphics[width=28.13pt,height=22.5pt]{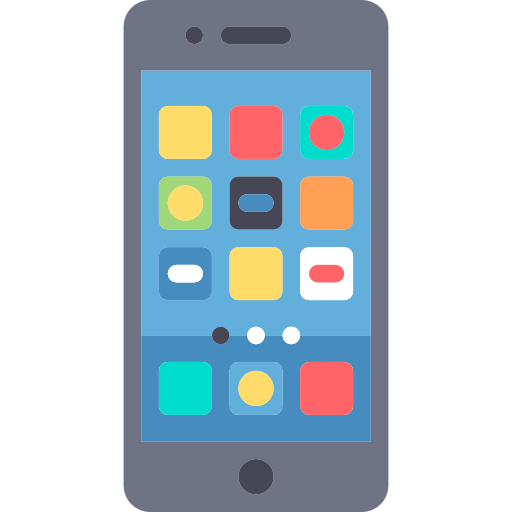}};
\draw  [dash pattern={on 4.5pt off 4.5pt}]  (219.5,153) -- (462.78,152.07) -- (494,152.28) ;
\draw [shift={(497,152.3)}, rotate = 180.39] [fill={rgb, 255:red, 0; green, 0; blue, 0 }  ][line width=0.08]  [draw opacity=0] (8.93,-4.29) -- (0,0) -- (8.93,4.29) -- cycle    ;
\draw    (221.5,128) -- (221.5,106) ;
\draw [shift={(221.5,104)}, rotate = 90] [color={rgb, 255:red, 0; green, 0; blue, 0 }  ][line width=0.75]    (10.93,-3.29) .. controls (6.95,-1.4) and (3.31,-0.3) .. (0,0) .. controls (3.31,0.3) and (6.95,1.4) .. (10.93,3.29)   ;
\draw    (221.5,128) -- (221.5,151) ;
\draw [shift={(221.5,153)}, rotate = 270] [color={rgb, 255:red, 0; green, 0; blue, 0 }  ][line width=0.75]    (10.93,-3.29) .. controls (6.95,-1.4) and (3.31,-0.3) .. (0,0) .. controls (3.31,0.3) and (6.95,1.4) .. (10.93,3.29)   ;
\draw  [draw opacity=0] (215.25,153) .. controls (215.25,150.65) and (217.15,148.75) .. (219.5,148.75) .. controls (221.85,148.75) and (223.75,150.65) .. (223.75,153) .. controls (223.75,155.35) and (221.85,157.25) .. (219.5,157.25) .. controls (217.15,157.25) and (215.25,155.35) .. (215.25,153) -- cycle ;
\draw (199,99.5) node  {\includegraphics[width=68.25pt,height=51.75pt]{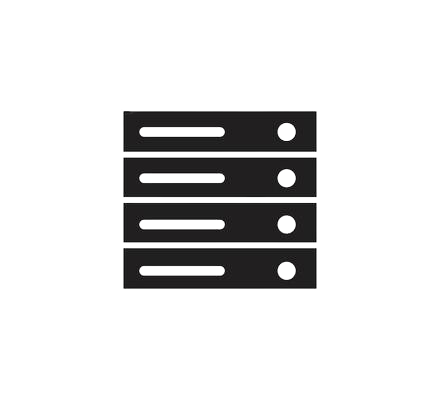}};
\draw  [fill={rgb, 255:red, 74; green, 74; blue, 74 }  ,fill opacity=1 ] (370,91) -- (385.5,91) -- (385.5,106.5) -- (370,106.5) -- cycle ;
\draw (234.75,204) node  {\includegraphics[width=28.13pt,height=22.5pt]{phone_icon.png}};
\draw  (194,281.8) -- (258,281.8)(200.4,235) -- (200.4,287) (251,276.8) -- (258,281.8) -- (251,286.8) (195.4,242) -- (200.4,235) -- (205.4,242)  ;
\draw    (209,281) -- (209,246) ;
\draw [shift={(209,243)}, rotate = 90] [fill={rgb, 255:red, 0; green, 0; blue, 0 }  ][line width=0.08]  [draw opacity=0] (8.93,-4.29) -- (0,0) -- (8.93,4.29) -- cycle    ;
\draw    (222,282) -- (222,258) ;
\draw [shift={(222,255)}, rotate = 90] [fill={rgb, 255:red, 0; green, 0; blue, 0 }  ][line width=0.08]  [draw opacity=0] (8.93,-4.29) -- (0,0) -- (8.93,4.29) -- cycle    ;
\draw   (212,296.75) -- (220,289) -- (228,296.75) -- (224,296.75) -- (224,312.25) -- (228,312.25) -- (220,320) -- (212,312.25) -- (216,312.25) -- (216,296.75) -- cycle ;
\draw  (185,368.8) -- (280,368.8)(194.5,322) -- (194.5,374) (273,363.8) -- (280,368.8) -- (273,373.8) (189.5,329) -- (194.5,322) -- (199.5,329)  ;
\draw    (195,344) .. controls (235,314) and (198,363) .. (238,333) ;
\draw    (238,333) .. controls (278,303) and (236,367) .. (276,337) ;
\draw  (365,280.8) -- (429,280.8)(371.4,234) -- (371.4,286) (422,275.8) -- (429,280.8) -- (422,285.8) (366.4,241) -- (371.4,234) -- (376.4,241)  ;
\draw    (380,280) -- (380,245) ;
\draw [shift={(380,242)}, rotate = 90] [fill={rgb, 255:red, 0; green, 0; blue, 0 }  ][line width=0.08]  [draw opacity=0] (8.93,-4.29) -- (0,0) -- (8.93,4.29) -- cycle    ;
\draw    (414,281) -- (414,257) ;
\draw [shift={(414,254)}, rotate = 90] [fill={rgb, 255:red, 0; green, 0; blue, 0 }  ][line width=0.08]  [draw opacity=0] (8.93,-4.29) -- (0,0) -- (8.93,4.29) -- cycle    ;
\draw   (383,295.75) -- (391,288) -- (399,295.75) -- (395,295.75) -- (395,311.25) -- (399,311.25) -- (391,319) -- (383,311.25) -- (387,311.25) -- (387,295.75) -- cycle ;
\draw  (356,367.8) -- (451,367.8)(365.5,321) -- (365.5,373) (444,362.8) -- (451,367.8) -- (444,372.8) (360.5,328) -- (365.5,321) -- (370.5,328)  ;
\draw    (367,343) .. controls (407,313) and (406,376) .. (446,346) ;
\draw  [dash pattern={on 0.84pt off 2.51pt}]  (277,104) -- (244.83,174.18) ;
\draw [shift={(244,176)}, rotate = 294.62] [color={rgb, 255:red, 0; green, 0; blue, 0 }  ][line width=0.75]    (10.93,-4.9) .. controls (6.95,-2.3) and (3.31,-0.67) .. (0,0) .. controls (3.31,0.67) and (6.95,2.3) .. (10.93,4.9)   ;
\draw  [dash pattern={on 0.84pt off 2.51pt}]  (370,106.5) -- (255.13,187.84) ;
\draw [shift={(253.5,189)}, rotate = 324.7] [color={rgb, 255:red, 0; green, 0; blue, 0 }  ][line width=0.75]    (10.93,-4.9) .. controls (6.95,-2.3) and (3.31,-0.67) .. (0,0) .. controls (3.31,0.67) and (6.95,2.3) .. (10.93,4.9)   ;
\draw  [dash pattern={on 0.84pt off 2.51pt}]  (285.5,105.5) -- (374.56,191.61) ;
\draw [shift={(376,193)}, rotate = 224.03] [color={rgb, 255:red, 0; green, 0; blue, 0 }  ][line width=0.75]    (10.93,-4.9) .. controls (6.95,-2.3) and (3.31,-0.67) .. (0,0) .. controls (3.31,0.67) and (6.95,2.3) .. (10.93,4.9)   ;
\draw  [dash pattern={on 0.84pt off 2.51pt}]  (378,109) -- (390.66,183.03) ;
\draw [shift={(391,185)}, rotate = 260.29] [color={rgb, 255:red, 0; green, 0; blue, 0 }  ][line width=0.75]    (10.93,-4.9) .. controls (6.95,-2.3) and (3.31,-0.67) .. (0,0) .. controls (3.31,0.67) and (6.95,2.3) .. (10.93,4.9)   ;

\draw (163,142.99) node [anchor=north west][inner sep=0.75pt]    {$( 0,0,0)$};
\draw (490.2,132.17) node [anchor=north west][inner sep=0.75pt]    {$x$};
\draw (273,12.72) node [anchor=north west][inner sep=0.75pt]    {$y$};
\draw (387.5,109.9) node [anchor=north west][inner sep=0.75pt]    {$\boldsymbol{\psi }_{2}$};
\draw (206,118.37) node [anchor=north west][inner sep=0.75pt]    {$d$};
\draw (266.4,66.89) node [anchor=north west][inner sep=0.75pt]   [align=left] {PA};
\draw (155.5,66) node [anchor=north west][inner sep=0.75pt]   [align=left] {{\small BS}};
\draw (373.4,68.89) node [anchor=north west][inner sep=0.75pt]   [align=left] {PA};
\draw (272,108.9) node [anchor=north west][inner sep=0.75pt]    {$\boldsymbol{\psi }_{1}$};
\draw (439,375.4) node [anchor=north west][inner sep=0.75pt]    {$f$};
\draw (267,376.6) node [anchor=north west][inner sep=0.75pt]    {$f$};
\draw (316,320.4) node [anchor=north west][inner sep=0.75pt]    {$H_{2}( f)$};
\draw (143,318.4) node [anchor=north west][inner sep=0.75pt]    {$H_{1}( f)$};
\draw (148,251.4) node [anchor=north west][inner sep=0.75pt]    {$h_{1}( \tau )$};
\draw (323,245.4) node [anchor=north west][inner sep=0.75pt]    {$h_{2}( \tau )$};
\draw (429,286.4) node [anchor=north west][inner sep=0.75pt]    {$\tau $};
\draw (249,285.4) node [anchor=north west][inner sep=0.75pt]    {$\tau $};
\end{tikzpicture}
    \caption{Illustration of the frequency-selective downlink channel in a PA system.}
    \label{fig:system_model_ISI}
\end{figure}

In this context, Fig. \ref{fig:system_model_ISI} shows that a pair of fixed PAs creates a frequency-selective downlink. Each PA contributes a separate propagation path that first travels through the leaky waveguide and then radiates through free space. Thus, the overall impulse response of the $m$-th use, which is denoted by $h_m(\tau)$, has two distinct delay taps and the transfer function $H_m(f)$ exhibits the corresponding spectral ripples. Because the PAs are several meters apart, the spacing between these taps is significant. For example, consider one user served by two PAs that are 3 meters apart. The extra distance inside the waveguide results in a delay of about $3\mathrm{m}/(c/n_e)$, which is approximately $14$ nanoseconds, while the additional free-space segment adds $3\mathrm{m}/c$, or approximately $10$ nanoseconds. The overall excess delay is therefore on the order of $24$ ns. In a single-carrier system operating at $500$ MHz, the symbol duration is only $T = 1/B \approx2$ ns, thus each symbol overlaps with more than ten previous symbols. To cope with this severe ISI, a single-carrier approach adopted in existing studies must either extend the symbol duration, which reduces the usable bandwidth, or insert long guard intervals. However, both measures impose a strict limit on the achievable data rate. This simple calculation highlights why an explicit ISI-aware design is essential when multiple, well-spaced PAs serve a user in a conventional time-division multiple access (TDMA) scheme.

To handle the frequency-selective channel response in \eqref{eq:frequency_response}, we adopt an OFDMA framework. To design the OFDMA frame we first determine the maximum delay spread introduced by the spatially separated PAs. For user $m$, considering the propagation both in the waveguide and free space, we define the composite delay of the signal radiated by the $n$-th PA as $\tau_{m,n}= \frac{\lVert\boldsymbol{\psi}_m-\boldsymbol{\psi}_{n}^{P}\rVert}{\lambda f_c}+\frac{\lVert\boldsymbol{\psi}_{0}-\boldsymbol{\psi}_{n}^{P}\rVert}{\lambda_g f_c}$. The worst excess delay is then $T_{\max}= \max_{m}\bigl\{\tau_{m,\max}-\tau_{m,\min}\bigr\}$, 
where $\tau_{m,\max}$ and $\tau_{m,\min}$ are the largest and smallest elements of $\{\tau_{m,n}\}_{n=1}^{N}$ for user $m$, thus we set the cyclic-prefix length $T_{\mathrm{CP}}\ge T_{\max}$.  
Next, for user $m$ we evaluate the root-mean-square (RMS) of its $N$ delays, $\sigma_{\tau,m} = \sqrt{\frac{1}{N}\sum_{n=1}^{N}\bigl(\tau_{m,n}-\bar{\tau}_m\bigr)^{2}}$, where $\bar{\tau}_m$ is the $m$-th user's mean delay. Setting $\sigma_\tau = \max_{m}\{\sigma_{\tau,m}\}$ ensures that the design protects the most dispersive link. Using the standard rule of thumb $B_c\approx1/(5\sigma_{\tau})$ for the coherence bandwidth, we select a fast Fourier transform (FFT) window that preserves orthogonality, i.e., $T_{\mathrm{FFT}} = T_{CP} + 1/B_c$.
With $T_{\mathrm{FFT}}$ fixed, the number of subcarriers is selected as the next power of two that fits in the available bandwidth, $K = 2^{\bigl\lceil \log_{2}\!\left(B\, T_{\mathrm{FFT}}\right)\bigr\rceil}$, facilitating radix-two FFT processing. The subcarrier spacing follows as \(\Delta f = B/K\), and the frequency of tone $k\in\{-K/2,\dots,K/2-1\}$ is $f_k = f_c + k\Delta f$. Removing the CP results in an efficiency loss quantified by $\eta=T_{\mathrm{FFT}}/(T_{\mathrm{FFT}}+T_{\mathrm{CP}})$. The aforementioned analysis applies directly to a single-user OFDM-based PA system.

For user $m$ on subcarrier $k$, the complex frequency-response sample is written as $H_{m,k} = H_m(f_k)$. Furthermore, the subcarrier-to-user mapping is captured by the binary indicator $b_{m,k}\in\{0,1\}$, where $b_{m,k}=1$ if and only if subcarrier $k$ is assigned to user $m$. Therefore, conventional OFDMA orthogonality enforces $\sum_{m=1}^{M}b_{m,k} = 1$ to ensure that every subcarrier is assigned to exactly one user, although a given user may occupy multiple subcarriers. When $b_{m,k}=1$, we allocate a non-negative power $p_{m,k}$ and the aggregate transmit power budget is written as 
\begin{equation}
    \begin{aligned}
        \sum_{m=1}^{M}\sum_{k=1}^{K}p_{m,k}\leq P_t,
    \end{aligned}
\end{equation}
with the implicit convention that $p_{m,k}=0$ when $b_{m,k}=0$.

With the above notation, the instantaneous downlink throughput of the user $m$ in the proposed OFDMA-based PA system is
\begin{equation}
    \begin{aligned}
        R_m^{(0)}(\boldsymbol{b}, \boldsymbol{p}) = \eta \Delta f \sum_{k=-K/2}^{K/2-1}b_{m,k} \log_{2}\left(1+\frac{\left|H_{m,k}\right|^2p_{m,k}}{N\Delta fN_0}\right),
    \end{aligned}
\end{equation}
where $\boldsymbol{b} = \{b_{m,k}\}$ and $\boldsymbol{p} = \{p_{m,k}\}$
collect the binary subcarrier assignments and the corresponding power loads, respectively. The additional factor $N$ in the denominator of the signal-to-noise ratio arises because the BS spreads the power $P_{m,k}$ uniformly across the $N$ PAs.

To guarantee balanced quality of service among users, we adopt a max-min design: the BS selects the subcarrier assignment matrix $\boldsymbol{b}$ and the power-loading matrix $\boldsymbol{p}$ so that the worst user achievable rate is maximized, while respecting both OFDMA orthogonality and the total transmit power budget. The optimization problem can be written as
\begin{equation*}\tag{\textbf{P1}}\label{eq:maxmin_with_theta}
\begin{aligned}
    \begin{array}{cl}
    \mathop{\mathrm{max}}\limits_{\boldsymbol{b}, \boldsymbol{p}} &\mathop{\mathrm{min}} \limits_{m}  R_{m}^{(0)}(\boldsymbol{b},\boldsymbol{p})\\
    \text{\textbf{s.t.}}
    & \mathrm{C}_1: \sum_{m=1}^{M}b_{m,k}=1,\\
    & \mathrm{C}_2:  p_{m,k}\geq0,\\
    & \mathrm{C}_3: \sum_{m=1}^{M}\sum_{k=1}^{K} p_{m,k} \leq P_t, \\
    & \mathrm{C}_4: b_{m,k}\in \{0,1\},
    \end{array}
\end{aligned}
\end{equation*}
  
Since the binary assignment variables $b_{m,k}$ appear in the logarithmic rate expression, the resulting formulation is a mixed-integer non-linear program (MINLP), which is NP-hard. We solve this using the proposed two-stage heuristic algorithm, outlined in Algorithm \ref{alg:greedy_WF}. First, a greedy subcarrier assignment produces a feasible matrix $\boldsymbol{b}$, and then, a per-user water-filling determines the power matrix $\boldsymbol{p}$, achieving near-optimal max-min fairness with polynomial complexity in the number of subcarriers.
Specifically, in the first stage, the binary assignment matrix \(\mathbf{b}\) is constructed by initializing each user’s provisional rate $R_{m}^{(0)} = 0$ and defining $\mathcal{K}$ as the set of unallocated subcarriers.  At each iteration, the user $m^{\star} = \arg\min_{m} R_m$ with the lowest current rate is selected.  For each subcarrier $k \in \mathcal{K}$, the metric $\Gamma_{m^{\star},k} = \frac{\lvert H_{m^{\star},k}\rvert^2}{\max_{q \neq m^{\star}} \lvert H_{q,k}\rvert^2}$ is computed to quantify the relative channel advantage.  The subcarrier $k^{\star} = \arg\max_{k\in \mathcal{K}} \Gamma_{m^{\star},k}$ is then allocated to user $m^{\star}$ by setting $b_{m^{\star},k^{\star}} = 1$ and removing $k^{\star}$ from $\mathcal{K}$.  The provisional rate of user \(m^{\star}\) is updated as $R_{m^{\star}} \;\leftarrow\; R_{m^{\star}}
+ \eta \,\Delta f \,
  \log_{2}\!\Bigl(1 + \tfrac{\lvert H_{m^{\star},k^{\star}}\rvert^2 \, P_t/K}
                       {N_0\,\Delta f\,N}\Bigr)$,
assuming equal power \(P_t/K\) per subcarrier.  This process repeats until $\mathcal{K}=\emptyset$, ensuring that each tone is assigned exactly once while progressively increasing the minimum provisional rate.
In the second stage, with $\mathbf{b}$ fixed, each user $m$ is allotted a power budget $P_t/M$ and performs classical water-filling across its assigned subcarriers $\{k : b_{m,k}=1\}$.  This power-loading step harnesses the channel’s frequency selectivity to maximize each user’s rate under the max–min fairness constraint.



\begin{algorithm}[h]
\caption{Two-stage max-min OFDMA resource allocation}
\label{alg:greedy_WF}
\small
\begin{algorithmic}[1]
  \Require Channel gains $G_{m,k}$, total power budget $P_{t}$
  \Ensure  Subcarrier assignment $\mathbf{b}$, power allocation $\mathbf{p}$

  \State $\mathbf{b} \gets \mathbf{0}$, $\mathbf{p} \gets \mathbf{0}$, $\mathcal{K}\gets\{-K/2,\dots,K/2-1\}$
  \State $R_m \gets 0,\ \forall m$  

  \While{$\mathcal{K}\neq\varnothing$}                       \Comment{\textsc{Stage 1: Greedy assignment}}
      \State $m^\star \gets \arg\min_m R_m$                  
      \State $k^\star \gets \displaystyle
             \arg\max_{k\in\mathcal{K}}
             \frac{\left|H_{m^\star,k}\right|^2}{\max_{q\neq m^\star}\left|H_{q,k}\right|^2}$
      \State $b_{m^\star,k^\star}\gets 1$;\quad
             $\mathcal{K}\gets\mathcal{K}\setminus\{k^\star\}$
      \State $R_{m^\star}\gets R_{m^\star}+
          \eta\,\Delta f\log_2\!\bigl(1+\tfrac{\left|H_{m^\star,k^\star}\right|^2P_t}{N_0\Delta f NK}\bigr)$
  \EndWhile

  \For{$m=1$ \textbf{to} $M$}                               \Comment{\textsc{Stage 2: Per-user water-filling}}
      \State $\mathcal{K}_m \gets\{k\,|\,b_{m,k}=1\}$        
      \State $P_m \gets P_t/M$                              
      \State Allocate $\{p_{m,k}\}_{k\in\mathcal{K}_m}$ by classical water-filling
  \EndFor
\end{algorithmic}
\end{algorithm}

\section{Simulation Results}
This section evaluates the performance of the proposed OFDMA-based PA system. For consistency, the system parameters are chosen as in \cite{sotiris_panos}, where the noise power is $-90$ dBm, the carrier frequency is $28$ GHz, the effective refractive index is $n_e = 1.4$, and the room area is $D_x \times D_y = 30\times10$ $\text{m}^2$. We compare the proposed scheme with two benchmarks. The first involves a single PA placed in the center of the room. In this setup, there is no ISI, since each user is served by a single PA and a single carrier in each time slot. The achievable rate for the $m$-th user is given by \cite{mmse_sc_fde}
\begin{equation}\label{eq:single_pinch_rate}
    \begin{aligned}
        R_{m}^{(1)} = \zeta^*_mB\log_{2}\left( 1 + \frac{\alpha_m|h_m|^2P_t}{N_0B}\right),
    \end{aligned}
\end{equation}
where $\zeta_m^*$ is the portion of time allocated to the $m$-th user, and the sum over all users satisfies $\sum_{m=1}^M\zeta_m^*=1$. It should be noted that $\zeta_m^*$ is optimally selected to maximize the minimum value of $R_m$, resulting in equal rates among users.

We also compare the proposed system with a second benchmark that uses $N$ uniformly deployed PAs and employs a TDMA scheme based on single-carrier frequency domain equalization. This approach utilizes a minimum mean square error equalizer to mitigate ISI, and the achievable rate is given by \cite{mmse_sc_fde}
\begin{equation}
    \begin{aligned}
        R_{m}^{(2)} = \zeta_m^* \eta B \log_{2}\left( 1 + \left(\frac{K}{\sum_{k=1}^{K}\frac{1}{\gamma_k+ 1}} - 1\right)\right),
    \end{aligned}
\end{equation}
where $\gamma_k = \frac{|H_{k,m}|^2P_t}{NKN_0\Delta f}$.

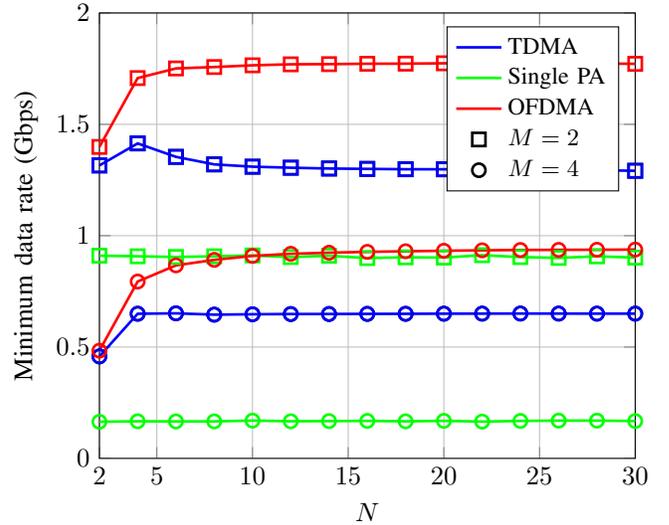
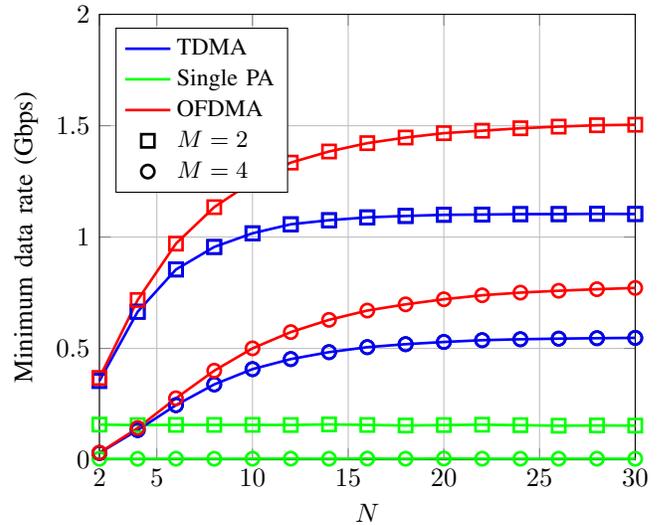
\begin{figure}[h]
    \centering
\begin{minipage}{.48\textwidth}
    \begin{tikzpicture}
        \begin{axis}[
            width=\linewidth,
            xlabel = $N$,
            ylabel = Minimum data rate (Gbps),
            xmin = 2,
            xmax = 30,
            ymin = 0,
            ymax = 2,
            xtick = {2,5,10,15,20,25,30},
            grid = major,
            legend cell align = {left},
            legend pos = north east,
            legend style={font=\small}
            ]
            \addplot[
            color= blue,
            no marks,
            line width = 1pt,
		      style = solid,
            mark=square*,
            mark repeat = 1,
            mark size = 2.5,
            ]
            table {figures/TDMA2opt_N2_5.000000e-02.txt};
            \addlegendentry{TDMA}
            
            \addplot[
            color= green,
            no marks,
            line width = 1pt,
		      style = solid,
            mark=square,
            mark repeat = 1,
            mark size = 2.5,
            ]
            table {figures/TDMA1opt_N2_5.000000e-02.txt};
            \addlegendentry{Single PA}
            
            \addplot[
            color= red,
            no marks,
            line width = 1pt,
		      style = solid,
            mark=square,
            mark repeat = 1,
            mark size = 2.5,
            ]
            table {figures/HGWF_N2_5.000000e-02.txt};
            \addlegendentry{OFDMA}

            \addplot[
            color= black,
            only marks,
            line width = 1pt,
		      style = solid,
            mark=square,
            mark repeat = 1,
            mark size = 2.5,
            ]
            table {figures/TDMA2opt_N2_5.000000e-02.txt};
            \addlegendentry{$M=2$}

            \addplot[
            color= black,
            only marks,
            line width = 1pt,
		      style = solid,
            mark=o,
            mark repeat = 1,
            mark size = 2.5,
            ]
            table {figures/TDMA2opt_N4_5.000000e-02.txt};
            \addlegendentry{$M=4$}


            \addplot[
            color= green,
            only marks,
            line width = 1pt,
		      style = solid,
            mark=square,
            mark repeat = 1,
            mark size = 2.5,
            ]
            table {figures/TDMA1opt_N2_5.000000e-02.txt};
            \addplot[
            color= green,
            only marks,
            line width = 1pt,
		      style = solid,
            mark=o,
            mark repeat = 1,
            mark size = 2.5,
            ]
            table {figures/TDMA1opt_N4_5.000000e-02.txt};

            \addplot[
            color= green,
            no marks,
            line width = 1pt,
		      style = solid,
            mark=o,
            mark repeat = 1,
            mark size = 2.5,
            ]
            table {figures/TDMA1opt_N4_5.000000e-02.txt};
            
            

            \addplot[
            color= blue,
            no marks,
            line width = 1pt,
		      style = solid,
            mark=o,
            mark repeat = 1,
            mark size = 2.5,
            ]
            table {figures/TDMA2opt_N4_5.000000e-02.txt};
            \addplot[
            color= blue,
            only marks,
            line width = 1pt,
		      style = solid,
            mark=o,
            mark repeat = 1,
            mark size = 2.5,
            ]
            table {figures/TDMA2opt_N4_5.000000e-02.txt};

            \addplot[
            color= blue,
            only marks,
            line width = 1pt,
		      style = solid,
            mark=square,
            mark repeat = 1,
            mark size = 2.5,
            ]
            table {figures/TDMA2opt_N2_5.000000e-02.txt};

            \addplot[
            color= red,
            only marks,
            line width = 1pt,
		      style = solid,
            mark=square,
            mark repeat = 1,
            mark size = 2.5,
            ]
            table {figures/HGWF_N2_5.000000e-02.txt};
          
            \addplot[
            color= red,
            no marks,
            line width = 1pt,
		      style = red,
            mark=square,
            mark repeat = 1,
            mark size = 2.5,
            ]
            table {figures/HGWF_N4_5.000000e-02.txt};

            \addplot[
            color= red,
            only marks,
            line width = 1pt,
		      style = red,
            mark=o,
            mark repeat = 1,
            mark size = 2.5,
            ]
            table {figures/HGWF_N4_5.000000e-02.txt};
            
            
            
        \end{axis}
    \end{tikzpicture}
\subcaption{$\beta = 0.05$}
\end{minipage}

\begin{minipage}{.48\textwidth}
    \begin{tikzpicture}
        \begin{axis}[
            width=\linewidth,
            xlabel = $N$,
            ylabel = Minimum data rate (Gbps),
            xmin = 2,
            xmax = 30,
            ymin = 0,
            ymax = 2,
            xtick = {2,5,10,15,20,25,30},
            grid = major,
            legend cell align = {left},
            legend pos = north west,
            legend style={font=\small}
            ]
            \addplot[
            color= blue,
            no marks,
            line width = 1pt,
		      style = solid,
            mark=square*,
            mark repeat = 1,
            mark size = 2.5,
            ]
            table {figures/TDMA2opt_N2_1.500000e-01.txt};
            \addlegendentry{TDMA}
            
            \addplot[
            color= green,
            no marks,
            line width = 1pt,
		      style = solid,
            mark=square,
            mark repeat = 1,
            mark size = 2.5,
            ]
            table {figures/TDMA1opt_N2_1.500000e-01.txt};
            \addlegendentry{Single PA}
            
            \addplot[
            color= red,
            no marks,
            line width = 1pt,
		      style = solid,
            mark=square,
            mark repeat = 1,
            mark size = 2.5,
            ]
            table {figures/HGWF_N2_1.500000e-01.txt};
            \addlegendentry{OFDMA}

            \addplot[
            color= black,
            only marks,
            line width = 1pt,
		      style = solid,
            mark=square,
            mark repeat = 1,
            mark size = 2.5,
            ]
            table {figures/TDMA2opt_N2_1.500000e-01.txt};
            \addlegendentry{$M=2$}

            \addplot[
            color= black,
            only marks,
            line width = 1pt,
		      style = solid,
            mark=o,
            mark repeat = 1,
            mark size = 2.5,
            ]
            table {figures/TDMA2opt_N4_1.500000e-01.txt};
            \addlegendentry{$M=4$}


            \addplot[
            color= green,
            only marks,
            line width = 1pt,
		      style = solid,
            mark=square,
            mark repeat = 1,
            mark size = 2.5,
            ]
            table {figures/TDMA1opt_N2_1.500000e-01.txt};
            \addplot[
            color= green,
            only marks,
            line width = 1pt,
		      style = solid,
            mark=o,
            mark repeat = 1,
            mark size = 2.5,
            ]
            table {figures/TDMA1opt_N4_1.500000e-01.txt};

            \addplot[
            color= green,
            no marks,
            line width = 1pt,
		      style = solid,
            mark=o,
            mark repeat = 1,
            mark size = 2.5,
            ]
            table {figures/TDMA1opt_N4_1.500000e-01.txt};
            

        
            \addplot[
            color= blue,
            no marks,
            line width = 1pt,
		      style = solid,
            mark=o,
            mark repeat = 1,
            mark size = 2.5,
            ]
            table {figures/TDMA2opt_N4_1.500000e-01.txt};
            \addplot[
            color= blue,
            only marks,
            line width = 1pt,
		      style = solid,
            mark=o,
            mark repeat = 1,
            mark size = 2.5,
            ]
            table {figures/TDMA2opt_N4_1.500000e-01.txt};

            \addplot[
            color= blue,
            only marks,
            line width = 1pt,
		      style = solid,
            mark=square,
            mark repeat = 1,
            mark size = 2.5,
            ]
            table {figures/TDMA2opt_N2_1.500000e-01.txt};

            \addplot[
            color= red,
            only marks,
            line width = 1pt,
		      style = solid,
            mark=square,
            mark repeat = 1,
            mark size = 2.5,
            ]
            table {figures/HGWF_N2_1.500000e-01.txt};
          
            \addplot[
            color= red,
            no marks,
            line width = 1pt,
		      style = red,
            mark=square,
            mark repeat = 1,
            mark size = 2.5,
            ]
            table {figures/HGWF_N4_1.500000e-01.txt};

            \addplot[
            color= red,
            only marks,
            line width = 1pt,
		      style = red,
            mark=o,
            mark repeat = 1,
            mark size = 2.5,
            ]
            table {figures/HGWF_N4_1.500000e-01.txt};
            
            
            
        \end{axis}
    \end{tikzpicture}
\subcaption{$\beta = 0.15$}
\end{minipage}
\caption{Minimum data rate versus number of PA for various users.}
\label{fig:min_rate_versus_pinches}
\end{figure}
Fig. \ref{fig:min_rate_versus_pinches} illustrates how the minimum data rate evolves with the number of PAs for two different user scenarios, under both low and moderate blockage conditions. It can be observed that the proposed OFDMA-based PA system consistently achieves higher minimum rates than both the TDMA scheme with SC-FDE and the single PA configuration, across all values of $N$ and for both $M=2$ and $M=4$. This confirms the robustness of OFDMA to variations in network density and blockage severity. The gap between OFDMA and TDMA becomes especially pronounced as $N$ increases. This trend highlights the limitations of the TDMA scheme, in which the equalizer experiences degraded effective SNR due to accumulated ISI caused by multiple delayed paths introduced by spatially separated PAs. In contrast, the OFDMA scheme benefits from the frequency selectivity of the channel. Certain subcarriers experience constructive interference, which directly contributes to higher achievable rates, as seen in the upward slope of the OFDMA curves. Furthermore, the figure illustrates how the flexibility of OFDMA enhances system performance. In our scheme, we can optimize both subcarrier allocation among users and power distribution per subcarrier, which is not possible in the TDMA baseline. This additional degree of control is reflected in the fact that even at higher values of $\beta$, the minimum rate under OFDMA remains significantly higher than both benchmarks. Finally, the results also show that the single PA suffers the most across all scenarios. As seen in Figs. \ref{fig:min_rate_versus_pinches}a and \ref{fig:min_rate_versus_pinches}b, the lack of spatial and frequency diversity limits its ability to cope with blockage, especially when $\beta = 0.15$, at which point its performance almost flattens. In contrast, the proposed system leverages multiple antennas and subcarriers to maintain strong performance regardless of the propagation environment.

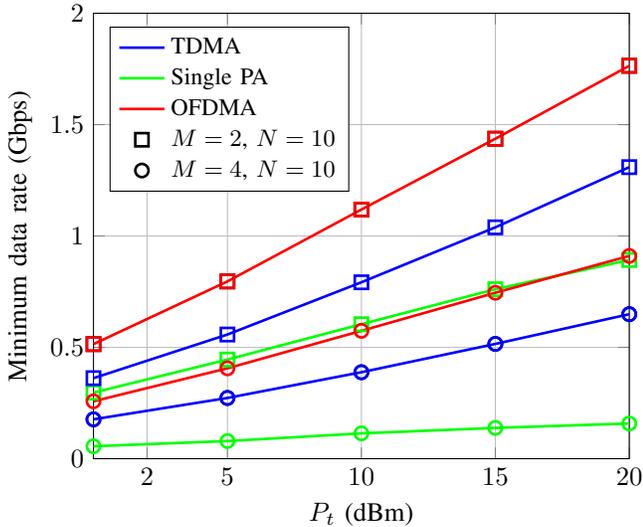
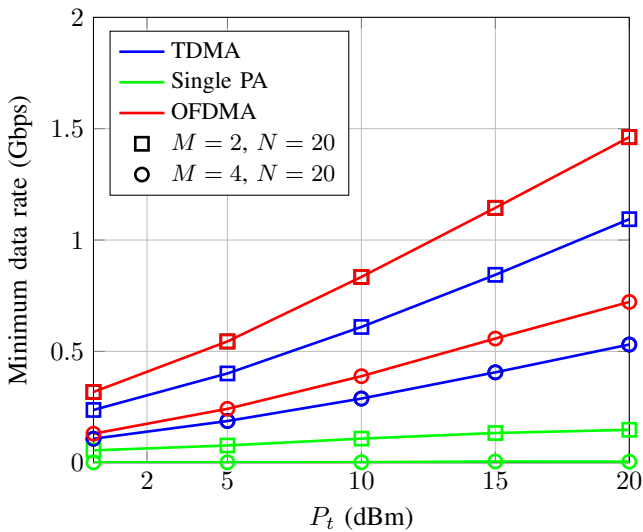
\begin{figure}[h]
    \centering
\begin{minipage}{.48\textwidth}
    \begin{tikzpicture}
        \begin{axis}[
            width=\linewidth,
            xlabel = $P_t$ (dBm),
            ylabel = Minimum data rate (Gbps),
            xmin = 0,
            xmax = 20,
            ymin = 0,
            ymax = 2,
            xtick = {2,5,10,15,20,25,30},
            grid = major,
            legend cell align = {left},
            legend pos = north west,
            legend style={font=\small}
            ]
            \addplot[
            color= blue,
            no marks,
            line width = 1pt,
		      style = solid,
            mark=square*,
            mark repeat = 1,
            mark size = 2.5,
            ]
            table {figures/TDMA2opt_N2_L10_5.000000e-02.txt};
            \addlegendentry{TDMA}
            
            \addplot[
            color= green,
            no marks,
            line width = 1pt,
		      style = solid,
            mark=square,
            mark repeat = 1,
            mark size = 2.5,
            ]
            table {figures/TDMA1opt_N2_L10_5.000000e-02.txt};
            \addlegendentry{Single PA}
            
            \addplot[
            color= red,
            no marks,
            line width = 1pt,
		      style = solid,
            mark=square,
            mark repeat = 1,
            mark size = 2.5,
            ]
            table {figures/HGWF_N2_L10_5.000000e-02.txt};
            \addlegendentry{OFDMA}

            \addplot[
            color= black,
            only marks,
            line width = 1pt,
		      style = solid,
            mark=square,
            mark repeat = 1,
            mark size = 2.5,
            ]
            table {figures/HGWF_N2_L10_5.000000e-02.txt};
            \addlegendentry{$M=2$, $N=10$}

            \addplot[
            color= black,
            only marks,
            line width = 1pt,
		      style = solid,
            mark=o,
            mark repeat = 1,
            mark size = 2.5,
            ]
            table {figures/TDMA2opt_N4_L10_5.000000e-02.txt};
            \addlegendentry{$M=4$, $N=10$}

            \addplot[
            color= green,
            only marks,
            line width = 1pt,
		      style = solid,
            mark=square,
            mark repeat = 1,
            mark size = 2.5,
            ]
            table {figures/TDMA1opt_N2_L10_5.000000e-02.txt};
            
            \addplot[
            color= green,
            only marks,
            line width = 1pt,
		      style = solid,
            mark=o,
            mark repeat = 1,
            mark size = 2.5,
            ]
            table {figures/TDMA1opt_N4_L10_5.000000e-02.txt};

            \addplot[
            color= green,
            no marks,
            line width = 1pt,
		      style = solid,
            mark=o,
            mark repeat = 1,
            mark size = 2.5,
            ]
            table {figures/TDMA1opt_N4_L10_5.000000e-02.txt};

            \addplot[
            color= blue,
            no marks,
            line width = 1pt,
		style = solid,
            mark=o,
            mark repeat = 1,
            mark size = 2.5,
            ]
            table {figures/TDMA2opt_N4_L10_5.000000e-02.txt};
            \addplot[
            color= blue,
            only marks,
            line width = 1pt,
		      style = solid,
            mark=o,
            mark repeat = 1,
            mark size = 2.5,
            ]
            table {figures/TDMA2opt_N4_L10_5.000000e-02.txt};

            \addplot[
            color= blue,
            only marks,
            line width = 1pt,
		      style = solid,
            mark=square,
            mark repeat = 1,
            mark size = 2.5,
            ]
            table {figures/TDMA2opt_N2_L10_5.000000e-02.txt};

            \addplot[
            color= red,
            only marks,
            line width = 1pt,
		      style = solid,
            mark=square,
            mark repeat = 1,
            mark size = 2.5,
            ]
            table {figures/HGWF_N2_L10_5.000000e-02.txt};
          
            \addplot[
            color= red,
            no marks,
            line width = 1pt,
		      style = red,
            mark=square,
            mark repeat = 1,
            mark size = 2.5,
            ]
            table {figures/HGWF_N4_L10_5.000000e-02.txt};

            \addplot[
            color= red,
            only marks,
            line width = 1pt,
		      style = red,
            mark=o,
            mark repeat = 1,
            mark size = 2.5,
            ]
            table {figures/HGWF_N4_L10_5.000000e-02.txt};     
        \end{axis}
    \end{tikzpicture}
\subcaption{$\beta = 0.05$}
\end{minipage}

\begin{minipage}{.48\textwidth}
    \begin{tikzpicture}
        \begin{axis}[
            width=\linewidth,
            xlabel = $P_t$ (dBm),
            ylabel = Minimum data rate (Gbps),
            xmin = 0,
            xmax = 20,
            ymin = 0,
            ymax = 2,
            xtick = {2,5,10,15,20,25,30},
            grid = major,
            legend cell align = {left},
            legend pos = north west,
            legend style={font=\small}
            ]
            \addplot[
            color= blue,
            no marks,
            line width = 1pt,
		      style = solid,
            mark=square*,
            mark repeat = 1,
            mark size = 2.5,
            ]
            table {figures/TDMA2opt_N2_L20_1.500000e-01.txt};
            \addlegendentry{TDMA}
            
            \addplot[
            color= green,
            no marks,
            line width = 1pt,
		      style = solid,
            mark=square,
            mark repeat = 1,
            mark size = 2.5,
            ]
            table {figures/TDMA1opt_N2_L20_1.500000e-01.txt};
            \addlegendentry{Single PA}
            
            \addplot[
            color= red,
            no marks,
            line width = 1pt,
		      style = solid,
            mark=square,
            mark repeat = 1,
            mark size = 2.5,
            ]
            table {figures/HGWF_N2_L20_1.500000e-01.txt};
            \addlegendentry{OFDMA}

            \addplot[
            color= black,
            only marks,
            line width = 1pt,
		      style = solid,
            mark=square,
            mark repeat = 1,
            mark size = 2.5,
            ]
            table {figures/HGWF_N2_L20_1.500000e-01.txt};
            \addlegendentry{$M=2$, $N=20$}

            \addplot[
            color= black,
            only marks,
            line width = 1pt,
		      style = solid,
            mark=o,
            mark repeat = 1,
            mark size = 2.5,
            ]
            table {figures/TDMA2opt_N4_L20_1.500000e-01.txt};
            \addlegendentry{$M=4$, $N=20$}

            \addplot[
            color= green,
            only marks,
            line width = 1pt,
		      style = solid,
            mark=square,
            mark repeat = 1,
            mark size = 2.5,
            ]
            table {figures/TDMA1opt_N2_L20_1.500000e-01.txt};
            
            \addplot[
            color= green,
            only marks,
            line width = 1pt,
		      style = solid,
            mark=o,
            mark repeat = 1,
            mark size = 2.5,
            ]
            table {figures/TDMA1opt_N4_L20_1.500000e-01.txt};

            \addplot[
            color= green,
            no marks,
            line width = 1pt,
		      style = solid,
            mark=o,
            mark repeat = 1,
            mark size = 2.5,
            ]
            table {figures/TDMA1opt_N4_L20_1.500000e-01.txt};

            \addplot[
            color= blue,
            no marks,
            line width = 1pt,
		style = solid,
            mark=o,
            mark repeat = 1,
            mark size = 2.5,
            ]
            table {figures/TDMA2opt_N4_L20_1.500000e-01.txt};
            \addplot[
            color= blue,
            only marks,
            line width = 1pt,
		      style = solid,
            mark=o,
            mark repeat = 1,
            mark size = 2.5,
            ]
            table {figures/TDMA2opt_N4_L20_1.500000e-01.txt};

            \addplot[
            color= blue,
            only marks,
            line width = 1pt,
		      style = solid,
            mark=square,
            mark repeat = 1,
            mark size = 2.5,
            ]
            table {figures/TDMA2opt_N2_L20_1.500000e-01.txt};

            \addplot[
            color= red,
            only marks,
            line width = 1pt,
		      style = solid,
            mark=square,
            mark repeat = 1,
            mark size = 2.5,
            ]
            table {figures/HGWF_N2_L20_1.500000e-01.txt};
          
            \addplot[
            color= red,
            no marks,
            line width = 1pt,
		      style = red,
            mark=square,
            mark repeat = 1,
            mark size = 2.5,
            ]
            table {figures/HGWF_N4_L20_1.500000e-01.txt};

            \addplot[
            color= red,
            only marks,
            line width = 1pt,
		      style = red,
            mark=o,
            mark repeat = 1,
            mark size = 2.5,
            ]
            table {figures/HGWF_N4_L20_1.500000e-01.txt};
        \end{axis}
    \end{tikzpicture}
\subcaption{$\beta = 0.15$}
\end{minipage}
\caption{Minimum data rate versus transmit power for various users.}
\label{fig:min_rate_versus_snr}
\end{figure}

Fig. \ref{fig:min_rate_versus_snr} presents the minimum data rate as a function of the transmit power for selected configurations of user and antenna numbers, under two blockage scenarios. The results show that systems with fewer users benefit more from increased transmit power, as indicated by the steeper slopes of their corresponding curves. In denser networks with more users, the slope becomes flatter, indicating that the performance improvement from higher transmit power is less pronounced. The OFDMA-based PA system consistently delivers the highest minimum data rate across all setups. This confirms its effectiveness in environments where ISI is introduced by the spatial separation of PAs. Notably, the proposed scheme achieves significantly higher worst-user rates, demonstrating its ability to enable reliable and fair communication in practical deployments.

\vspace{-1em}
\section{Conclusion}
In this work, we examined a practical, low-complexity PA deployment in which the PAs are uniformly placed along the waveguide to deliver LoS service to multiple users without the need for continual tracking or reconfiguration. We demonstrated that this deployment naturally creates a strongly frequency-selective channel, which introduces intrinsic ISI. Utilizing this behavior, we proposed an OFDMA-based transmission framework that turns the resulting frequency diversity to advantage while suppressing ISI. To achieve user fairness, we formulated a max-min resource-allocation problem and solved it with a lightweight two-stage algorithm-greedy subcarrier assignment followed by per-user water-filling, which maintains polynomial complexity. Simulation results under realistic mmWave propagation and blockage conditions demonstrated substantial gains in the minimum achievable rate over time-division and single-carrier benchmarks, establishing OFDMA as an effective means of scaling PA systems to dense, high-capacity indoor networks. In addition, future work will explore how optimizing the pre-placement of the PAs jointly with the OFDMA resource allocation developed here can provide new design guidelines and unlock further gains for large indoor PA deployments.
\bibliographystyle{IEEEtran}
\bibliography{Bibliography}

\end{document}